\documentclass[a4paper,12pt]{article}
\usepackage{amssymb}
\usepackage{amsmath}
\usepackage{eurofont}
\newcommand{\beq}{\begin{equation}}
\newcommand{\eeq}{\end{equation}}

\newcommand{\p}{\mathbf p}
\begin{document}
\title{\textbf{Inconsistency of the judgment matrix in the AHP method and the decision maker's knowledge}}
\author{Anna Szczypi\'nska\\ Institute of Physics, University of Silesia, \\ Uniwersytecka
4, Pl 40007 Katowice, Poland \\ e-mail:
zanna12@wp.pl\\
Edward W.~Piotrowski\\ Institute of Mathematics,
University of Bia\l ystok,\\ Lipowa 41, Pl 15424 Bia\l ystok,
Poland\\ e-mail: ep@alpha.uwb.edu.pl
}
\date{}
\maketitle

\begin{abstract}
In this paper we propose a method for a quantitative estimation of the decision maker's knowledge in the context of the Analytic Hierarchy Process (AHP) in cases, where the judgment matrix is inconsistent. We show that the matrix of deviation from the transitivity condition corresponds to the rate matrix for transaction costs in the financial market.  For the quantitative estimation of the decision maker's professionalism, we apply the Ising model and  thermodynamics tools.
\end{abstract}


\section*{Introduction}
Do we know what people take into consideration when they make decisions? Do they make the same decision in the same circumstances each time? How does the decision maker's knowledge influence the decision that (s)he makes? These are very interesting questions in the theory of decision making. Some people make rational decisions, which are determined by their knowledge about the exterior conditions. Their decisions are easy to predict. On the other hand, others, who are guided by emotions or very subjective opinions, make completely different choices in the same situations. Their behaviour is irrational and unclear for other people. Prof. R. Aumann, a mathematician and the Nobel Memorial Prize winner for game theory in economics, explains that human life is composed of infinitely repeated games in which we take part \cite{aumann}. The result of every previous game influences the next one. What is rational and scientifically certain in one game, in another may harm. Taking this into consideration, people often do not make the decision that is the best for them at a given moment in other people's opinion, but the one  that will bring better results in the future. People very often react quite disproportionally to what has happened, because their reaction is proportional to what might happen if some event was to be repeated. Similarly, we very often resign from something in one game, in order to avoid problems in another. Everything depends on the decision maker's goals and the knowledge (s)he possesses. These goals and his/her knowledge are very often inaccessible to other people. 

We consider the problem of the estimation of the decision maker's knowledge in the context of the Analytic Hierarchy Process (AHP) proposed by Saaty \cite{saatyahp} as a method of multi-criterion decision making. AHP involves decomposing a complex decision into a hierarchy of goals, criteria, sub-criteria and alternatives comparing the properties of each possible pair of elements at each level as a matrix and synthesising the priorities. The major problem with this method is to find weights which order the objects and reflect the recorded judgments in situations where the judgments are not transitive, that is, the judgment matrix is inconsistent. As we argue in the paper \cite{pmmahp}, human thinking is intransitive, because e.g. new knowledge requires that we change our minds. Moreover, modern decision makers must take into consideration more and more alternatives. In such situations, conservation consistency of the judgment matrix is impossible \cite{7b,15bz}. In the paper \cite{pmmahp} we identify the matrix of deviation from the transitivity condition as the rate matrix for ``transaction costs'', which expresses decision maker's aversions and preferences. We propose the method of estimation his/her professionalism, which takes into account these ``costs''. The concept of transaction costs does not appear accidentally here, because, as we show, they correspond to the margin, with whom we deal e.g. in exchange offices. 

An estimation of the decision maker's professionalism is fundamental with reference to the credibility and the quality of their decisions. The AHP method is widely applied in such domains as: prioritisation, resource allocation, public policy, strategic planning and many more \cite{2b,gw,lgv}, and the qualitative but not quantitative nature of the decision agents makes a simple estimation of a concrete choice impossible. For the quantity estimation of the decision maker's effectiveness we apply the Ising model \cite{stat}. The Ising Hamiltonian determines  ``profit'' from  a decision strategy. Applying the thermodynamics approach \cite{thermod}, we describe classes of decision makers that achieve the same profits in different ways. Moreover, we determine the effective algorithm of decision making, which leads to the maximum profit. This allows us to answer the question, what is the most effective way of a decision maker's behaviour with fixed market parameters and no zero transaction costs. The determined strategy corresponds to the clairvoyant strategy. Then, we show that with the help of the thermodynamics parameter---the temperature, which is the measure of achievable profit, we can compare the achievements of people who solve completely different decision problems with different transaction costs and at different times. This parameter allows us to compare an abstract decision maker's profit with the gain of the investor as well. Finally, we propose a market interpretation of the discrete version of the Fisher information \cite{fisher} in the case where the probability density function takes the value $0$ or $1$. 

\section{Transaction costs}
To define and describe transaction costs, we observe the mechanism of its formation. Let us consider the prices of currencies in exchange offices. We will not be able to buy and sell  currency at the same price. The exchange offices make money by selling the currency to customers for more than they paid to buy it and by buying the currency from customers for less than they will receive when they sell it. The difference between these prices is called a spread. We try to find a formula which determines the value of the spread. Let us consider the quotations of two currencies EUR and USD in an exchange office. Let $\ln b$ be the logarithm of the bid price of EUR/USD and $\ln a$ be the logarithm of the ask price. Then, $\ln \frac{1}{a}$ is the logarithmic exchange rate for converting US dollars to euros. From the principle of insufficient reason, see \cite{dupont}, if there is no reason to prefer one exchange rate over another, then it is sensible to assume a priori that both transactions of the exchange -- euros for dollars and dollars for euros can occur with equal probability  $\frac{1}{2}$.  According to this principle we define the commission $\epsilon$ as the expectation value of the logarithmic exchange rates
\begin{equation}
\epsilon := \frac{\ln b+\ln \frac{1}{a}}{2}=\frac{\ln b-\ln a}{2}\,.\label{tc} 
\end{equation}
We consider the logarithmic exchange rates because only in logarithms can the product of rates resulting from a series of transactions in an exchange office have the property of additivity. 

\section{Geometry of the market}
Now we will describe the market in which the decision processes take place. Let us consider a market of goods and/or obligations, criteria of judgments, information and so forth, which we denote by $G$. If we assume the infinite divisibility of goods, then $G$ has the structure of an $N$-dimensional linear space over the reals \cite{gofm}. Elements of this space are called baskets. For any basket $\p\in G$ we have a unique decomposition into the goods which make it~up $$ \p=\sum_{\mu=1}^N p_{\mu} \mathsf{g}_\mu\,, $$ in some fixed basis of normalised unit goods $(\mathsf{g}_1,\dots,\mathsf{g}_N)$. The element $\mathsf{g}_\mu \in G$ is the $\mu$th market goods and the coefficient $p_\mu \negthinspace\negthinspace\in \negthinspace\mathbb{R}$ is called the $\mu$th coordinate of the basket and expresses its quantity. The market quotation $U$ is a linear map $U(\mathsf{g}_\nu,~\cdot~): G \rightarrow {\mathbb R}$ which assigns to every basket $p$ its current value in units of $\mathsf{g}_\nu$:
$$
\label{pipi-rzut} (U \p)_\nu = U(\mathsf{g}_\nu,\p)  = \sum_{\mu=1}^N
U(\mathsf{g}_\nu,\mathsf{g}_\mu) p_\mu\, ,
$$
where $U(\mathsf{g}_\nu,\mathsf{g}_\mu)$ is the relative price of the unit of $\mu$th asset given in units of $\nu$th asset. The elements $u_{\nu\mu}:=U(\mathsf{g}_\nu,\mathsf{g}_\mu)$, for $\nu,\mu =1,\dots,N$, make the $N\negthinspace \times\negthinspace N$ matrix of market rates $\mathbf{U}$.\\

In the AHP method $\mathbf{U}=(u_{\nu\mu})$ is a pairwise comparison matrix of judgments. AHP is based on several steps \cite{saatyahp}. Firstly, a complex decision is decomposed into a hierarchy of goals, criteria, sub-criteria and alternatives, which influence the solution to the problem. Then, a decision maker expresses his/her own local preferences by comparisons of the properties of each possible pair of elements at each level. The pairwise comparisons are organised into a square matrix of the relative judgments. The next step is to compute a vector of priorities from this matrix that indicates the relative importance of each alternative. Alternatives can be quantitative (goods) or qualitative (personal preferences). The decision makers express their own opinions in the pairwise comparison of alternatives in kilos, metres, euros for the quantity-based judgments, or as equal, strong, very strong and so on for the quality-based judgments. In the second case, these must be converted into numbers, see \cite{6b}. The problem is to find weights which order the objects and reflect the recorded judgments in a case where the pairwise comparison matrices are inconsistent, that is, the judgments are intransitive $u_{\nu\rho}u_{\rho\mu}\neq u_{\nu\mu}$, see \cite{pmmahp}. In the paper \cite{pmmahp} we consider the intransitive case of the AHP model for which the matrix of market rates was not antisymmetrical, that is $u_{\mu\nu}\neq \frac{1}{u_{\nu\mu}}$. Such a deviation of opinion from the condition of transitivity expresses decision maker's aversions and preferences and corresponds to the transaction costs in the financial market. Let us note that we can identify the logarithm of the entry of the judgment matrix with the logarithmic exchange rate for converting one good to another. In the case of the EUR/USD, the logarithmic exchange rate for converting euros to dollars equals $\ln u_{\text{\$}\text{\euro}}=\ln b$ and dollars to euros---$\ln u_{\text{\euro}\text{\$}}=\ln \frac{1}{a}$. From Eq.~(\ref{tc}) it follows that if we take the transaction costs into consideration, we obtain the following formulas
$$\ln u_{\text{\$} \text{\euro}}=\ln \frac{1}{u_{\text{\euro} \text{\$}}}+2\epsilon\,\;\;\;\;\;\;\;\text{and}\,\;\;\;\;\;\;\;\ln u_{\text{\euro} \text{\$}}=\ln \frac{1}{u_{\text{\$} \text{\euro}}}+2\epsilon \,.$$
As we can see, thanks to the commission defined like (\ref{tc}) any of the exchange rates for converting euros to dollars and dollars to euros is not more preferred. Both of these rates are equally profitable for a broker and (s)he has no reason to prefer one to the other. Hence, we can express all logarithmic exchange rates for converting the $\mu$th good to the $\nu$th good in the AHP method as 
$$\ln u_{\nu\mu}=\ln \frac{1}{u_{\mu\nu}}+2\epsilon_{\nu\mu}\,,$$
for $\nu, \mu=1,\ldots,N$, where $\epsilon_{\nu\mu}$ is a suitable commission. Let us note that the elements $\epsilon_{\nu\mu}$ have the property of antisymmetry. 

\section{Measurement of the decision maker's knowledge taking into account the transaction costs---the Ising model}\label{pomiarpref}
We want to measure the decision maker's preferences in the AHP model in the context of the profit that (s)he achieves. In order to speak of profit in the case of decision problems, it is necessary that the decision maker selects some asset $\mathsf{g}_{0}$ that will be the ``currency'' and expresses the judgments in the form of their relative prices $u_{0\mu}=U(\mathsf{g}_0 ,\mathsf{g}_\mu)$ in relation to this asset, see \cite{pmmahp,gofm}. Of course, the decision maker's preferences may change in time, but (s)he can always express them in respect to the fixed ``currency''. Let $u_{01t},\ldots ,u_{0Nt}$ be the relative prices of all criterions at the moment $t$ in relation to $\mathsf{g}_{0t}$. Let us consider a strategy that consists of choosing only one criterion at the successive moments $t=1,2,\ldots ,k$. If the decision maker takes the $\mu$th criterion at the moment $t$, then $$h_{\mu t}:=\ln \frac{u_{0\mu t}}{u_{0\mu (t-1)}}$$ denotes the logarithmic rate of return from this criterion in the time interval $[t\negthinspace-\negthinspace1,t]$. But, if we assume that at the moment $t\negthinspace-\negthinspace1$ the person who makes decision had the $\nu$th criterion, we have to take into consideration the transaction cost $\epsilon_{\mu \nu}$ from converting the $\nu$th criterion into the $\mu$th criterion. Let the transaction costs be fixed. Let $n_{\mu t}$ denotes the sentence \textit{the decision maker chooses the $\mu$th criterion at the moment t}. Then, $[n_{\mu t}]$ is the Iverson bracket and it takes the value $1$ if \textit{the decision maker chooses the $\mu$th criterion at the moment t} or $0$ otherwise. For every scenario of the decision maker's behaviours we can give the sequence $([n_{\mu t}])$ whose elements fulfil the formula 
\begin{equation}
\sum_{\mu =1}^{N}[n_{\mu t}]=1\,,\label{suma}
\end{equation}
for $t\negthinspace=\negthinspace1,2,\ldots ,k$. Let us note that the cost $\epsilon_{\mu \nu}$ appears if and only if (s)he chooses the $\nu$th criterion at the moment $t\negthinspace-\negthinspace1$ and converts it to the $\mu$th criterion at the moment $t$. We describe this cost in the interval $[0,k]$ with the help of propositional calculus. Then, we can express the profit of the person who makes a decision in the interval $[0,k]$ by the formula
$$
 H([n_{1 1}],\ldots ,[n_{N k}])\negthinspace:=\negthinspace\negthinspace\sum_{\mu ,\,t}h_{\mu t}[n_{\mu t}]-\negthinspace\sum_{\nu ,\,\mu ,\,t}\negthinspace \epsilon_{\mu \nu} [ \neg (n_{\mu t}\negthinspace\Rightarrow\negthinspace n_{\mu (t-1)}\negthinspace)\wedge \neg (n_{\nu (t-1)}\negthinspace\Rightarrow\negthinspace n_{\nu t}\negthinspace)].
$$
The sequence $([n_{\mu t}])$, for $\mu\negthinspace=\negthinspace1,\ldots ,N$;\, $t\negthinspace=\negthinspace1,2,\ldots ,k$, defines the decision maker's strategy and $([n_{1 1}],\ldots ,[n_{N k}])\negthinspace:=\negthinspace([n_{1 1}],\ldots ,\negthinspace[n_{N 1}],\ldots ,\negthinspace[n_{1 k}],\ldots ,\negthinspace[n_{N k}])$ describes the pure strategy. 

Using the Boolean lattice on the logical values of the received sentence forms, we obtain $[\neg x]:=1-[x]$, $[x\wedge y]:=[x][y]$ and we note that 
$[x][x]=[x]$, for $[x]\in \left\lbrace 0,1\right\rbrace$. Then, from Eq.~(\ref{suma}) and the rules of propositional calculus, we obtain the following formula for the decision maker's profit
\begin{equation}
 H([n_{1 1}],\ldots ,[n_{N k}])=\sum_{\mu ,\,t}h_{\mu \,t}[n_{\mu t}] -\sum_{\nu ,\,\mu ,\,t} \epsilon_{\mu \nu}[n_{\nu (t-1)}] [n_{\mu t}]\,.\label{isi}
\end{equation}
If we assume periodic boundary conditions, that is, $[n_{\mu \,0}]=[n_{\mu \,k}]$,  Eq.~(\ref{isi}) represents a Hamiltonian of an Ising chain \cite{stat}. If we rewrite this equation in the terms of spins $S_{\mu t}:=[n_{\mu t}]-\frac{1}{2}$, we obtain
\begin{align}
H(S_{1 1},\ldots ,S_{N k}) & =\sum_{\mu ,\,t}h_{\mu \,t}\,S_{\mu t}-\sum_{\nu ,\,\mu ,\,t}\epsilon_{\mu \nu}S_{\mu t} -\sum_{\nu ,\,\mu ,\,t} \epsilon_{\mu \nu}S_{\nu (t-1)}S_{\mu t}-\notag \\
& -\sum_{\nu ,\,\mu ,\,t} \frac{1}{4}\epsilon_{\mu \nu}+\sum_{\mu ,\,t}\frac{1}{2}h_{\mu \,t}\,,\notag
\end{align}
that is, the energy of the Ising model for particles with the spin $\frac{1}{2}$ in the alternating magnetic field $h_{\mu t}- \epsilon_{\nu \mu}$ modified about some constants.\\

We see that, because of the ability of the Ising model to represent the main features of the critical behaviour of many different systems, it is very popular not only in physics, but also in econophysics and economics \cite{ball}. In 1974 H.~F\"{o}llmer introduced interactions to microeconomics and, for the first time, created a model of economy that was based on rules similar to those of the Ising model for magnets \cite{folmer}.

\section{The thermodynamics of the canonical ensemble}
We know that some people make rational decisions that are easy to predict. On the other hand, others make irrational and unclear choices in other people's opinions. We cannot predict such behaviour, but we can measure their indefiniteness and its influence on the quality of decision making measured by profit, as formulated in Eq.~(\ref{isi}). To measure the indefiniteness or confusion of the person who makes decisions, we have to compare his/her profit with the outcomes of other decision makers. Therefore, we classify the decision makers according to the profit which they achieve, see \cite{arbitr}. Let us consider all mixed strategies that for a given price sequence $(h_{\mu t})$ bring the same profit as equivalent. These strategies can be parametrised by $2^{N k}$ weights $p_{[n_{1 1}],\ldots ,[n_{N k}]}$ corresponding to the pure strategies. This set of strategies is called the canonical statistical ensemble \cite{isihara, thermod}. In the canonical statistical ensemble there is the  decision maker who makes only rational decisions and achieves the profit that describes the ensemble. Let us choose the strategy that maximises the entropy to represent the class of decision makers. We will call such a strategy the canonical strategy. We do not know anything about the person who applies it except the expectation value of the profit $E(H)$, which (s)he achieves. In order to find the weights of the canonical strategy, we have to calculate a conditional extreme of the Boltzmann-Shannon entropy \nolinebreak\cite{shannon}
\begin{equation}
S_{\left( p_{[n_{1 1}],\ldots ,[n_{N k}]}\right) }:=-E(\ln
p_{[n_{1 1}],\ldots ,[n_{N k}]}) \label{entropy}
\end{equation}
subject to the constraints:
$$
 \sum_{[n_{1 1}],\ldots ,[n_{N k}]=0}^{1}\negthinspace\negthinspace p_{[n_{1 1}],\ldots ,[n_{N k}]}=1\, \eqno(7\text{a})
$$
and
$$
E(H([n_{1 1}],\ldots ,[n_{N k}]))=\text{const}\,. \eqno(7\text{b})
$$
We can interpret the condition $(7\text{b})$ as the weakened balance condition. It follows from the Newton's third law of motion, which in the financial markets implies the non-existence of the free lunch phenomenon. This problem was discussed in the paper \cite{deterministic}. In order to find the weights of the canonical strategy, we use the standard method of the Lagrange multipliers, that is, we solve the following differential equation
$$
\mathrm{d} S_{\left( p_{[n_{1 1}],\ldots ,[n_{N k}]}\right) }-\beta\, \mathrm{d}E(H([n_{1 1}],\ldots ,[n_{N k}]))-\,\zeta\,\mathrm{d}\negthinspace\negthinspace\negthinspace\negthinspace
\sum_{[n_{1 1}],\ldots ,[n_{N k}]=0}^{1}\negthinspace\negthinspace p_{[n_{1 1}],\ldots ,[n_{N k}]}=\nopagebreak
$$
$$
-\negthinspace\negthinspace\sum_{[n_{1 1}],\ldots ,[n_{N k}]=0}^{1}
\negthinspace\negthinspace\bigl(\,\ln p_{[n_{1 1}],\ldots,[n_{N k}]} +1
+\beta\,H([n_{1 1}],\ldots ,[n_{N k}]) +\zeta\bigr)\,\mathrm{d}p_{[n_{1 1}],\ldots,[n_{N k}]}=0\,,
$$
where $\beta$ and $\zeta$ are the Lagrange multipliers. The above equation must be fulfilled for all values $\mathrm{d}p_{[n_{1 1}],\ldots,[n_{N k}]}$, therefore 
$$\ln p_{[n_{1 1}],\ldots,[n_{N k}]} +1 +\beta \,H([n_{1 1}],\ldots ,[n_{N k}]) +\zeta=0\,.$$ 
From this equation we find that the weights of the canonical strategy are equal
\begin{equation}
p_{[n_{1 1}],\ldots,[n_{N k}]}={\mathrm e}^{-\beta\,H([n_{1 1}],\ldots ,[n_{N k}])-\zeta-1}\,.\label{wagi}
\end{equation}
Now, we can eliminate the Lagrange multiplier $\zeta$. The weights are normalised, see Eq.~$(7\text{a})$, therefore 
$$\sum_{[n_{1 1}],\ldots ,[n_{N k}]=0}^{1}\negthinspace\negthinspace p_{[n_{1 1}],\ldots ,[n_{N k}]}=\sum_{[n_{1 1}],\ldots ,[n_{N k}]=0}^{1}\negthinspace\negthinspace {\mathrm e}^{-\beta\,H([n_{1 1}],\ldots ,[n_{N k}])-\zeta-1}=1$$
and
$${\mathrm e}^{-\zeta-1}=\frac{1}{\sum\limits_{\phantom{a}[n_{1 1}],\ldots ,[n_{N k}]=0}^{1}\negthinspace\negthinspace {\mathrm e}^{-\beta\,H([n_{1 1}],\ldots ,[n_{N k}])}}\,.$$ After substituting this formula to (\ref{wagi}), we obtain the Gibbs distribution function \cite{isihara} 
$$ p_{[n_{11}],\ldots,[n_{Nk}]}:=\frac{{\mathrm e}^{-\beta\,H([n_{1 1}],\ldots ,[n_{N k}])}}{
\sum\limits_{\phantom{a}[n_{11}],\ldots,[n_{Nk}]=0}^{1}\negthinspace\negthinspace {\mathrm e}^{-\beta\,H([n_{1 1}],\ldots ,[n_{N k}])}}\,,
$$
which expresses the dependence of the weights of the canonical strategy on the profit $H([n_{1 1}],\ldots ,[n_{N k}])$, which is the result from the pure strategies. The function
\begin{align}
Z &:=\negthinspace\negthinspace\sum\limits_{\phantom{a}[n_{11}],\ldots,[n_{Nk}]=0}^{1}
\negthinspace\negthinspace{\mathrm e}^{-\beta\, H([n_{1 1}],\ldots ,[n_{N k}])}=\notag \\ & =\negthinspace\negthinspace\sum\limits_{\phantom{a}[n_{11}],\ldots,[n_{Nk}]=0}^{1}
\negthinspace\negthinspace{\mathrm e}^{-\beta\,\left(\frac{1}{2}\sum\limits_{\mu ,\,t}h_{\mu t}[n_{\mu t}]+\frac{1}{2}\sum\limits_{\nu ,\,t}h_{\nu (t-1)}[n_{\nu (t-1)}] -\negthinspace\sum\limits_{\nu ,\,\mu ,\,t}\negthinspace \epsilon_{\mu \nu}\,[n_{\nu (t-1)}] [n_{\mu t}]\right) }\notag
\end{align}
is called the partition function in thermodynamics and the inverse of the Lagrange multiplier $\beta$ -- the temperature of the canonical ensemble\footnote{What in physics denotes expressing temperature in the natural unit---the Boltzmann constant \cite{isihara}.},  $T=\frac{1}{\beta}$. In the canonical statistical ensemble the temperature is fixed, $T=\text{const}$. We can express the partition function $Z$, for the profit given by (\ref{isi}), with the help of the transition matrices that depend on the successive moments $t-1$ and $t$:
$$
M(t)_{[n_{\nu (t-1)}][n_{\mu t}]}:=\mathrm{e}^{-\beta\,\left(\frac{1}{2}h_{\mu t}[n_{\mu t}] +\frac{1}{2}h_{\nu (t-1)}[n_{\nu (t-1)}]- \epsilon_{\mu \nu}\,[n_{\nu (t-1)}][n_{\mu t}]\right)} \,.
$$
Thus, the matrix $M(t)$ is a $2\negthinspace\times\negthinspace 2$ matrix given by
\begin{displaymath}
M(t)=\left( \begin{array}{cc}
1 & \mathrm{e}^{-\frac{1}{2}\beta\, h_{\mu t}}\\
\mathrm{e}^{-\frac{1}{2}\beta\, h_{\nu (t-1)}} & \mathrm{e}^{-\beta\,(\frac{1}{2}h_{\mu t}+\frac{1}{2}h_{\nu (t-1)} -\epsilon_{\mu \nu})}
\end{array}\right). 
\end{displaymath}
Then, for convenient periodic boundary conditions, the function $Z$ can be written as the trace of the transition matrices product 
\begin{align}
Z & =\negthinspace\negthinspace\negthinspace\negthinspace\sum\limits_{\phantom{a}[n_{11}],\ldots,[n_{Nk}]=0}^{1} \negthinspace\negthinspace M(1)_{[n_{\mu k}][n_{\nu 1}]}M(2)_{[n_{\nu 1}][n_{\eta 2}]}\cdots
 M(k)_{[n_{\theta (k-1)}][n_{\mu k}]}\notag \\
&=\mathrm{Tr}\prod\limits_{t=1}^k M(t)\,.\label{trace}
\end{align}

Let us try to find the best strategy of the decision making that leads to the maximum expectation profit. Because the entries of the matrices $M(t)$ depend on time via $h_{\mu t}$, we cannot consider the proper value problem of the partition function because it does not lead to its dense form. That is why we use the ($\mathrm{max,\,+}$) algebra \cite{quadrat, thermod}.

\section{The ($\mathrm{max,\,+}$) algebra of extreme strategies}
From the definition of the entropy (\ref{entropy}) and since $p_{[n_{1 1}],\ldots ,[n_{N k}]}\negthinspace=\negthinspace\frac{1}{Z}\mathrm{e}^{-\beta H}$, we obtain
\begin{align}
S& =-E(\ln p_{[n_{1 1}],\ldots ,[n_{N k}]})=-\negthinspace\negthinspace\sum_{[n_{1 1}],\ldots ,[n_{N k}]=0}^{1}\negthinspace\negthinspace p_{[n_{1 1}],\ldots ,[n_{N k}]}\ln
p_{[n_{1 1}],\ldots ,[n_{N k}]}\notag \\
&=\ln Z+\beta E(H)\,.\notag 
\end{align}
After substituting $\beta =\frac{1}{T}$ to the above formula, we obtain the following relation among the entropy, the average profit and the partition function
\begin{equation}
T\ln Z +E(H)=TS\,.\label{zaleznosc}
\end{equation}
From the above equation the strategy giving the maximum profit can be found by calculating  the limit for $T\rightarrow 0^-$. Because, the entropy is a positive function and it attains its maximum value $\ln 2^{Nk}$ when all probabilities equal $\frac{1}{2^{Nk}}$, it is bounded from above $S\leq Nk \ln 2$. Therefore, the maximum expected profit $H_+$ is equal to
$$
H_+:=\lim\limits_{T\rightarrow0^{-}}E(H)=-\lim\limits_{T\rightarrow0^{-}}T\,\ln Z=
\lim\limits_{\beta\rightarrow -\infty}\log_{\mathrm{e}^{-\beta}}Z\,.
$$ 
If we take into consideration Eq.~(\ref{trace}), we obtain
$$
H_+=\lim\limits_{\beta\rightarrow -\infty}\log_{\mathrm{e}^{-\beta}}\negthinspace\left( \mathrm{Tr}\prod\limits_{t=1}^k M(t)\right)\,.
$$
Let us define the ($\mathrm{max,\,+}$) algebra that follows from the elementary properties of the logarithmic function: 
$$
\log_\varepsilon(\varepsilon^a\,\varepsilon^b)=a+b\,,\,\,\,\,\,\,\,\,
\lim\limits_{\varepsilon\rightarrow +\infty}\log_\varepsilon(\varepsilon^a+\varepsilon^b)=
\max(a,b)\,,
$$
so, instead of adding the matrix elements we use the operation $\mathrm{max}$ of taking the maximal element of them, and instead of their multiplication---their sum. Let 
\begin{align}
\tilde{M}(t)_{[n_{\nu (t-1)}][n_{\mu t}]}:& =\log_{\mathrm{e}^{-\beta}} M(t)_{[n_{\nu (t-1)}][n_{\mu t}]}\negthinspace\negthinspace\notag \\
& =\negthinspace \frac{1}{2}h_{\mu t}[n_{\mu t}]+\frac{1}{2}h_{\nu (t-1)}[n_{\nu (t-1)}]-\epsilon_{\mu \nu}\,[n_{\nu (t-1)}][n_{\mu t}]\notag
\end{align}
be the logarithm of the transition matrix, then the logarithm of the transition matrices product is defined as follows  
$$
\bigl(\tilde{M}(t)\times \tilde{M}(t\negthinspace+\negthinspace1)
\bigr)_{[n_{\nu (t-1)}][n_{\theta (t+1)}]}\negthinspace:=\max\limits_{[n_{\mu t}]}\bigl(
\tilde{M}(t)_{[n_{\nu (t-1)}][n_{\mu t}]}\negthinspace+\tilde{M}(t\negthinspace+\negthinspace
1)_{[n_{\mu t}][n_{\theta (t+1)}]}\bigr)\,. 
$$
Hence, the decision maker can reconstruct the sequence $([n_{11}],\ldots,[n_{Nk}])$, which corresponds to the optimal strategy of a clairvoyant leading to the maximal profit, by investigating the matrix elements contributing to the function
$$
\bigl(\tilde{M}(1)\times \tilde{M}(2)\times\cdots\times\tilde{M}(k)\bigr)_{[n_{\mu k}][n_{\mu k}]}:=
$$
$$=\max\limits_{[n_{11}],\ldots,[n_{Nk}]}\bigl(
\tilde{M}(1)_{[n_{\mu k}][n_{\nu 1}]}\negthinspace+\tilde{M}(2)_{[n_{\nu 1}][n_{\eta 2}]}+\cdots+\tilde{M}(k)_{[n_{\theta (k-1)}][n_{\mu k}]}\bigr)\,.
$$
The maximal element determines the maximum possible profit.

\section{Assessment of the decision maker's professionalism}
As we know, the entropy allows us to measure the qualities of a decision (or financial) expert. However, the temperature $T$ can be used to measure the financial gain and compare achievements on different markets, during different time frames, and with different transaction costs. Therefore, with the help of this parameter we can compare seemingly different phenomena, e.g. the decision maker's profit and the gain of the investor. 

In the previous section we considered the canonical statistical ensembles that were characterised by the same profit from the decision strategy. As the representative of the canonical ensemble we  chose the strategy that maximised the entropy. It is described by the Gibbs distribution function. Let us consider how the motion of the prices $h_{\mu t}$ influences the expectation value of the profit from such a strategy with the fixed parameters $n_{\mu t}$. Let $S$ be the independent variable. Keeping in mind the formulas that describe the profit $H$ and the partition function $Z$, we can calculate the differential of the entropy of the Gibbs distribution function from Eq.~(\ref{zaleznosc})
\begin{align}
T \mathrm{d}S & =T\frac{-\beta\negthinspace\negthinspace\negthinspace\negthinspace\negthinspace \sum\limits_{\phantom{a}[n_{11}],\ldots,[n_{Nk}]=0}^{1}
\negthinspace\negthinspace\negthinspace{\mathrm e}^{-\beta\left(\sum\limits_{\mu ,\,t}h_{\mu t}[n_{\mu t}] -\sum\limits_{\nu ,\,\mu ,\,t} \epsilon_{\mu \nu}[n_{\nu (t-1)}] [n_{\mu t}] \right)  }\negthinspace\sum\limits_{\mu ,\,t}[n_{\mu t}]\mathrm{d} h_{\mu t}}{Z}+\mathrm{d} E(H)\notag\\
& =-\sum\limits_{\mu ,\,t}E\left( [n_{\mu t}]\right)\mathrm{d} h_{\mu t} +\mathrm{d} E(H)=-\sum\limits_{\mu ,\,t} \overline{[n_{\mu t}]}\mathrm{d} h_{\mu t} +\mathrm{d} E(H)\,,\notag
\end{align}
where $\overline{[n_{\mu t}]}$ is the average value of the $\mu$th good at the moment $t$. From the above formula it follows that the temperature of the canonical strategy is equal 
$$
T=\frac{\partial E(H)}{\partial S}\,.
$$
Then, the temperature measures the intensity of the change of the expectation value of the profit $E(H)$ from the canonical strategy caused by its entropy change, see \cite{thermod}. The value $-T$ is the price of a unit of the entropy $S$, therefore, the bigger the absolute value of the temperature of the decision maker's canonical strategy is, the greater is his/her confusion. Infinite temperature has the canonical strategy representing monkey strategies. On the other hand, the higher the class of the specialist is, the smaller entropy and the absolute value of the temperature his/her canonical statistical ensemble has. The best results are achieved by the decision makers with a temperature close to zero.  

\section{Market interpretation of Fisher information}
In the previous sections we considered the Boltzmann-Shannon entropy as a measure of the decision maker's professionalism. It expresses his/her confusion. But there are other measures that we can employ to estimate the knowledge of the person who makes decisions, e.g. Fisher information \cite{fisher}. The Fisher information has very important applications in physics because it allows many physical laws and constants to be calculated, e.g. it explains the prescription for constructing Lagrangians, see \cite{sciencefrom}. The source of this information is the measure of the expected error that arises in the estimation process of  physical quantities. Let us assume that we want to estimate a parameter of the value $\theta$ on the basis of the imperfect observation $y=\theta + x$ of $\theta$, where $x$ is random noise. Let us consider the class of unbiased estimates obeying $E( \hat{\theta} )=\theta \,.$ Then, the mean-square error $e^2$ in the estimate $\hat{\theta}$ obeys the Cramer-Rao inequality \cite{rao}
\beq
e^2 I\geq 1\,,\label{raocramer}
\eeq
where $I$ is the Fisher information given by the formula  
\beq I=\int \frac{p'(x)}{p(x)}\mathrm{d}x \,,\;\;\;\;\;\;\;\;\;\;p':=\frac{\mathrm{d}p}{\mathrm{d}x}\,.\label{ifisher}
\eeq
The quantity $p(x)$ denotes the probability density function for the noise $x$. From Eq.~(\ref{raocramer}), it follows that the Fisher information is a quality metric of the estimation procedure, since, the quality increases ($e$ decreases) as I increases. Moreover, the Fisher information is a measure of system disorder like the Shannon entropy. The stronger the disorder is, the smaller the value of the Fisher information $I$ is \cite{sciencefrom}.

The Fisher information is a related quantity to the Shannon entropy, but it has different analytic properties. Whereas, the Shannon entropy is the global measure of the information about the system, the Fisher information is the local measure. Let us consider the discrete form of these quantities. We can express  an increment of the Shannon entropy as
$$
S=-\triangle x\sum_{\mu}p(x_\mu)\ln p(x_\mu)\,,\;\;\;\;\;\;\;\;\;\;\triangle x\rightarrow 0\,.
$$   
In the above formula the summation can be in any order and it does not depend on the organisation of the system. The value of $S$ remains constant, therefore entropy is said to be a global measure of the behaviour of $p(x_\mu)$. However, the discrete form of the Fisher information (\ref{ifisher}) is given by the formula \cite{sciencefrom}
\beq
I=\triangle x^{-1}\sum_{\mu}\frac{[p(x_{\mu+1})-p(x_\mu)]^2}{p(x_\mu)}\,.\label{dyskrfisher}
\eeq 
This formula is not symmetrical. After replacement of the variables  $x_{\mu+1}\leftrightarrow x_{\mu}$, we obtain the other value of $I$. Moreover, if the system undergoes a rearrangement, the local slope values  $[p(x_{\mu+1})-p(x_\mu)]/\triangle x$ will change drastically, so the sum (\ref{dyskrfisher}) will also change markedly. However, for $p(x_\mu)=0$ discontinuities of the Fisher information and $I$ goes toward infinity, since $I$ is thereby sensitive to local rearrangement of the system. Now, we try to generalise the formula (\ref{dyskrfisher}) so it expresses information about the $\mu$th criterion for the pure strategy defined in subsection \ref{pomiarpref}.

Let us consider such case of the Fisher information in which the probability distribution function $p(x_\mu)$ takes only two values $0$ and $1$. We can identify it with the Iverson bracket defined in subsection \ref{pomiarpref} for the $\mu$th criterion. Taking the property (\ref{suma}) into consideration, we can express the Fisher information about the $\mu$th criterion for the pure strategy as 
$$
I_\mu :=\sum_t \left( \texttt{if} \;\; [n_{\mu t}]=[n_{\mu (t+1)}]\;\; \texttt{then}\;\;0 \;\;\texttt{else}\;\; \frac{1}{2}\frac{([n_{\mu (t+1)}]-[n_{\mu t}])^2}{\frac{[n_{\mu (t+1)}]+[n_{\mu t}]}{2}}\right) \eqno{(11\text{a})}$$
$$\quad\quad=\sum_t \left( \texttt{if}\;\; [n_{\mu t}]=[n_{\mu (t+1)}]\;\;\texttt{then}\;\;0 \;\;\texttt{else}\;\;\frac{([n_{\mu (t+1)}]-[n_{\mu t}])^2}{[n_{\mu (t+1)}]+[n_{\mu t}]}\right).\eqno{(11\text{b})} 
$$
Let us note that the above formula is symmetrical; it does not have discontinuities points and it takes into consideration the local changes of the function $[n_{\mu t}]$. If no changes concerning the $\mu$th criterion happen at two successive moments, that is $[n_{\mu t}]=[n_{\mu (t+1)}]$, then its contribution to the sum ($11\text{b}$) in the interval $[t, t+1]$ is zero. Moreover, the formula ($11\text{b}$) has a natural market interpretation. With accuracy to a multiplicative constant the value $I_\mu$ corresponds to the sum of transaction costs that are connected with the exchange of the $\mu$th criterion for an arbitrary one from among the remaining $N\negthinspace-\negthinspace1$ criterions and an arbitrary criterion for the $\mu$th criterion. The factor $1/2$, which appears in Eq.~($11\text{a}$), corresponds to  half of the total transaction costs related to the exchange of the $\mu$th criterion. The latter half of these costs will be assigned to the criterion which follows the exchange or which is exchanged for the $\mu$th criterion.

The Fisher information from mutually isolated systems like Boltzmann-Shannon entropy has an additivity property  \cite{sciencefrom}. Hence, the information about the total costs from the strategy, which consists of choosing only one criterion from among $N$ criterions at successive moments, is the sum of the Fisher information concerning particular criterions
$$
I=\sum\limits_{\mu=1}^N I_\mu\,.
$$
We can interpret the Fisher information defined via transaction costs as the measure of the decision maker's professionalism. If we assume that the expected value of the profit from the strategy is fixed, then the decision makers whose accomplishments are greater will bear higher costs\footnote{e.g. in the form of taxes.}. These higher costs are placed on them in order to restrict their above-average profits and earnings at the same time. In the case of the confused experts the costs can be reduced because there are big chances to take advantage of the unfortunate moves. Hence, among decision makers belonging to the same canonical statistical ensemble, there are professionals who bear the high costs and laymen who pay low margins. Everybody achieves the same profits from their strategies. The question is, who we would choose as our adviser? 

\section{Conclusions}
We have proposed a method that allows us to measure the decision maker's knowledge quantitatively. The search for such methods is necessary in times, when a single decision may bring huge losses or huge gains. Thanks to the use of the models and tools of physics, we can obtain not only new qualitative information about the nature of decision problems, but also a wider look at the physics phenomena. The pricing of the decision alternatives and application of the statistical physics tools allows us study the phenomena about the qualitative nature numerically and notice connections between decisions and investment problems \cite{arbitr}. The clairvoyant strategy answers the question: what is the best way of a decision maker's behaviour taking into account intransitivity of his/her opinion with fixed market parameters? However, with the help of the temperature, we can compare the professionalism of people who solve decision problems in different fields, at different time, and with different transaction costs.

\section*{Acknowledgements}
This work was supported in part by the scientific network \textit{Laboratorium Fizycznych Podstaw Przetwarzania Informacji} sponsored by the \textbf{Polish Ministry of Science and Higher Education} (decision no 106/E-345/BWSN-0166/2008).

\end{document}